\documentstyle [12pt] {article}

\catcode`@=12
\begin{document}
\vspace{0.5in}
\oddsidemargin -.375in
\newcount\sectionnumber
\sectionnumber=0
\def\be{\begin{equation}}
\def\ee{\end{equation}}
\thispagestyle{empty}
\begin{flushright} UH-511-780-94\\January 1994\\
\end{flushright}
\vspace {.5in}
\begin{center}
{\Large\bf Corrections to mass scale predictions in SO(10)\\ GUT with
higher dimensional operators\\}
\vspace{.5in}
{\bf Alakabha Datta${}^{a)}$},
{\bf Sandip Pakvasa${}^{a)}$} and {\bf Utpal Sarkar${}^{b)}$\\}
\vspace{.1in}
${}^{a)}$ {\it
Physics Department, University of Hawaii at Manoa, 2505 Correa
Road, Honolulu, HI 96822, USA.}\\
${}^{b)}$ {\it Theory Group,
Physical Research Laboratory, Ahmedabad - 380009, India.}\\
\end{center}

\begin{abstract}

We calculate the two loop
contributions  to
 the predictions of the
mass scales in an SO(10) grand unified theory. We consider the
modified unification scale boundary conditions due to the
non-renormalizable higher dimensional terms arising from
quantum gravity or spontaneous compactification of extra
dimensions in Kaluza-Klein type theory.  We find  the range
of these couplings  which  allows left-right symmetry
to survive till very low energy (as low as $\sim$ TeV) and
still be compatible with the latest values of
$\sin^2 \theta_W$ and $\alpha_s$ derived from LEP. We consider
both the situation when the left-right parity is broken and
conserved.We consider both supersymmetric and non-supersymmertic versions
of the SO(10) theory.Taking the D-conserved non-susy case as an example
we calculate the effects of moderate threshold uncertainties at the heavy
scale, due to the unknown higgs masses,on the gravity induced couplings.
\end{abstract}
\newpage
\baselineskip 24pt

There are many extensions of the standard model, which are
suggested on various aesthetical grounds. But so far
experiments could not find anything
 which is not predicted by the standard model. In
other words, the standard model is consistent with all the
experiments carried out so far, although there are applealing
reasons to believe that there is physics beyond the standard
model. In the standard model the (V--A) nature of the theory is
put in by hand, whereas in an left-right symmetric
extension \cite{lr} of the standard model this comes about through
spontaneous symmetry breaking.

In the left-right symmetric extension of the standard model, at
higher energies the gauge group is extended to a left-right
symmetric group $G_{LR} \equiv SU(2)_L \otimes SU(2)_R \otimes
U(1)_{B-L}$. When appropriate higgs fields acquire vacuum
expectation value ($vev$), this group breaks down to one of its
subgroup $G_{std} \equiv SU(2)_L \otimes U(1)_Y$. There will
then be new scalar and gauge particles of mass of the order of
this symmetry breaking scale $M_R$. The mixing of these gauge
bosons with the standard model gauge bosons puts lower bound on
this scale.

The $K_L - K_S$ mass difference gives a lower bound \cite{klks}
of about 1.6  TeV on $M_R$ from the box diagram with both $W_L$
and $W_R$ exchanges. However, this constraint is subject to the
assumption of manifest left-right symmetry, which is to assume
that the Kobayashi-Maskawa matrices of the left- and the right-
handed sectors are same. In absence of this artificial symmetry
(which does not have any natural explanation) the bound
\cite{arc} on $M_R$ is relaxed to  300 GeV. From the
direct search \cite{cdf} at CDF the lower bound on $M_{W_R}$ is
520 GeV. This bound is not applicable to left-right symmetric
models where the $W_R$ couples only to the heavy neutrinos,
which again decay very fast. The strongest bound on $M_R$
comes from an analysis \cite{gb,alt} of the precision
measurement of the $Z-$pole from the CERN $e^+ e^-$ collider
LEP \cite{lep}. From a fit of the 1992 data and for the
commonly chosen higgs triplet fields for the left-right
symmetry breaking, the lower bound on $M_R$ is of the order of
TeV.

In the standard model the three gauge coupling constants are
free parameters and are all different.
 This has a natural expalnation in grand
unified theories \cite{gut} in which the strong and the
electroweak interaction are only low energy manifestations of a
single interaction. The GUT interaction is a gauge interaction
based on a simple gauge group with only one gauge coupling
constant. Through spontaneous symmetry breaking this breaks
down to a low energy symmetry group. Then the different coupling
constants evolve in different ways to give the present day low
energy coupling constants. Some of the attractive features of GUTS
were their natural explanation of the problem
of baryogenesis, and their unique prediction of proton decay.
However, proton decay has not yet been observed and the
question of baryogenesis took a completely different shape
following the observation of large anomalous baryon number
nonconservation at high temperatures in the presence of
sphaleron fields. The main interest in GUTs remains is its
unification of coupling constants and charge quantization.

Recently there has again been an upsurge of interest
\cite{amaldi,lrgut,palas,alak,des} in GUTs following the
precision measurement of the three gauge coupling constants at
LEP. The normalised gauge coupling constants for the groups
$SU(3)_c$, $SU(2)_L$ and $U(1)_Y$, as obtained \cite{lep} from
analyzing the LEP data, are given by,

\begin{eqnarray}
\alpha_1(M_z) & = & .16887 \pm .000040 \nonumber\\
\alpha_2(M_z) & = & .03322 \pm .00025 \nonumber\\
\alpha_3(M_z) & = & .120 \pm .007
\end{eqnarray}

\noindent respectively. With the minimal particle content, it
is not possible to unify all the three coupling constants at
any energy. This apparently rules out \cite{amaldi} minimal
$SU(5)$ GUT and any GUTs without any intermediate scales and
new particles unless the effect of gravity modifies the
situation.

It was further pointed out that the scale of the intermediate
symmetry breaking can be severely constrained by the present
values of the gauge coupling constants. For the minimal
supersymmetric GUTs, the supersymmetry breaking scale $M_R \;
\sim $ 1 TeV gives a good fit \cite{amaldi} to evolve all the
gauge couplings to an unification point. However, threshold
effects and higher order corrections make this scale uncertain
by orders of magnitude. This makes the threshold effects and
higher order corrections very important in studying the
evolution of the gauge coupling constants in the light of the
LEP data.

It was pointed out that if one studies any GUTs with left-right
symmetric group $G_{LR}$ as one of its intermediate symmetry
group, then the present LEP data severely constrain
\cite{lrgut} this symmetry breaking scale $M_R$. For any GUTs
and any number of new symmetries above $M_R$, one obtains a
lower bound $$M_R > 10^9 \:\:\: GeV.$$ This bound can be
relaxed \cite{chang} if one breaks the left-right parity and
the left-right symmetric group $G_{LR}$ at different scales.

If now signatures of the right handed gauge bosons are found in
the next generation accelerators (since the experimental lower
bound is only around a TeV), that will not however mean that
there is inconsistency in GUTs. It was shown that in a very
specific supersymmetric $SO(10)$ GUT one can satisfy \cite{des}
the unification constraitn with low $M_R$. The details of this
deserves further study.

Since the GUT scale is very close to the Planck scale, the
effects of gravity may not be negligible. It was shown that if
effects of gravity are considered through higher dimensional
operators, then even the minimal $SU(5)$ GUT with no new
particle content may be consistent \cite{patra} with the LEP
data and proton decay.

We have studied \cite{alak} the effect of gravity to see if the
constraints on $M_R$ can be relaxed. We considered higher
dimensional nonrenormalizable operators which may arise due to
quantum gravity or spontaneous compactification of extra
dimensions in the Kaluza-Klein type theory and their effect in
the $SO(10)$ lagrangian. The GUT scale boundary condition was
found to be modified and for certain choice of parameters low
$M_R$ could be made consistent with $SO(10)$ GUT. In this paper
we present details of our analysis. Here we include the
threshold effects and study the two loop evolution of the
coupling constants which are also very significant in these
analyses. First we present the formalism and then present our
analysis. At the end we summarize our results.

Higher dimensional operators were considered originally
\cite{hidim} to help solve some problems in fermion masses. The idea
is  to find out if
the low energy physics  contains some signatures of gravity
effects. In all these analysis the coupling constants in these
nonrenormalizable terms are free parameters. Someday we may learn
 if such coupling constants may
arise from gravity naturally.

In our analysis we consider dimension five and dimension six operators
when the contribution from dimension five operator vanish.We note that the
effect of all operators higher than dimension six can be  absorbed in
the couplings of the dimension six operators and hence their inclusion does
not increase the number of parameters.We therefore consider
only  dimension five
and dimension six operators in our analysis.

The main objective of our study is to look for consistency of
low $M_R$. For this purpose we consider the symmetry breaking
chain,

\begin{eqnarray}
SO(10) & {M_U \atop \longrightarrow}& SU(4) \times
SU(2)_L \times SU(2)_R \nonumber \\
&& \left[ \equiv G_{PS} \right] \nonumber \\
&& \nonumber \\
&{M_I \atop \longrightarrow}& SU(3)_c \times SU(2)_L \times
SU(2)_R \times U(1)_{(B-L)} \nonumber  \\
&& \left[ \equiv G_{LR} \right] \nonumber  \\
&& \nonumber\\
&{M_R \atop \longrightarrow}& SU(3)_c \times SU(2)_L  \times
U(1)_Y \nonumber  \\
&& \left[ \equiv G_{std} \right] \nonumber \\
&& \nonumber\\
&{M_W \atop \longrightarrow}& SU(3)_c \times U(1)_{em}.
\label{chain}
\end{eqnarray}

Near the scale $M_U \sim 10^{16}$ GeV or higher the gravity effects are
not negligible. But we assume that any theory beyond this scale
respects the $SO(10)$ symmetry. Then the lagrangian will
contain all the usual $SO(10)$ invariant dimension 4
interaction terms and in addition will contain $SO(10)$
invariant higher dimensional nonrenormalizable terms. These
higher dimensional terms will be suppressed by the Planck scale
(in theories \cite{hidim} where these terms are induced by
quantum gravity) or by the Kaluza--Klein compactification scale
(in theories \cite{kk} where these terms are induced by
spontaneous compactification of the extra dimensions in the
Kaluza-Klein type theories), which can even be two orders of
magnitude below the Planck scale.

The lagrangian can be written as,
\be
{L} = {L}_R + {L}_{NR}
\ee
where the first part of the lagrangian contains all
the renormalizable dimension 4 terms including the $SO(10)$
gauge invariant term,
\be
L=-{1 \over 2} {\rm Tr} (F_{\mu\nu} F^{\mu\nu})
\ee
where,
$$
F_{\mu \nu}=\partial_{\mu}A_{\nu} - \partial_{\nu}A_{\mu} -ig
\lbrack A_{\mu}, A_{\nu} \rbrack
$$ $$
A_{\mu}=A_{\mu}^i {\lambda_i \over 2} \nonumber
$$
with
$$
{\rm Tr}(\lambda_i \lambda_j)={1 \over 2} \delta_{ij}, \nonumber
$$
where, the $\lambda$'s are the $SO(10)$ generators. The
nonrenormalizable part of the lagrangian contains all the
higher dimensional $SO(10)$ invariant terms. We are presently
interested in only dimension 5 and 6 terms, which are given as,
$$
L=\sum_{n=5} L^{(n)} \label{eqn2}
$$
\begin{equation}
L^{(5)}=-{1 \over 2} {\eta^{(1)} \over M_{Pl}} {\rm Tr} (F_{\mu\nu}
\phi F^{\mu\nu})
\end{equation}
\begin{eqnarray}
L^{(6)}=&-{1 \over 2} {1 \over M_{Pl}^2} \biggl\lbrack
\eta_a^{(2)}\lbrace {\rm Tr} (F_{\mu\nu} \phi^2 F^{\mu\nu}) +
{\rm Tr} (F_{\mu\nu} \phi F^{\mu\nu}\phi)\rbrace + \nonumber \\
&\eta_b^{(2)} {\rm Tr}(\phi^2){\rm Tr} (F_{\mu\nu} F^{\mu\nu}) +
\eta_c^{(2)} {\rm Tr} (F^{\mu\nu}\phi) {\rm Tr} (F_{\mu\nu}
\phi) \biggr\rbrack
\end{eqnarray}
where $\eta^{(n)}$ are dimensional couplings of the higher
dimensional operators. When any higgs scalar $\phi$ acquires
$vev$ $\phi_0$, these operators induce effective dimension 4
terms, which modifies the boundary conditions at the scale
$\phi_0$.

Let us consider the symmetry breaking chain[\ref{chain}] at the
scale $M_U$. We shall first consider the case when this
symmetry breaking is mediated by the $vev$ of a 54-plet of
higgs. In this case the left-right parity is broken at $M_R$
only when $SU(2)_R$ is broken and the gauge coupling constants
$g_L$ and $g_R$ corresponding to the groups $SU(2)_L$ and
$SU(2)_R$ respectively evolve similarly between $M_U$ and $M_R$
so that $g_L(M_R) = g_R(M_R)$. In the second case we shall
consider the symmetry breaking at the scale $M_U$ by a 210-plet
of higgs. This breaks the discrete left-right parity symmetry
\cite{chang} D, so that $g_L$ and $g_R$ evolve in a different
way below $M_U$ and as a result one obtains $g_L(M_R) \neq
g_R(M_R)$.

In the D-conserving case, the symmetry breaking at $M_U$ takes
place when the 54-plet higgs $\Sigma$ of $SO(10)$ acquires a
$vev$,
\be
\langle \Sigma \rangle = \displaystyle
\frac{1}{\sqrt{30}}\:\: \Sigma_0 \:\: {\rm diag}(1, 1, 1, 1, 1,
1, -\frac{3}{2}, -\frac{3}{2}, -\frac{3}{2}, -\frac{3}{2}).
\ee
where, $\Sigma_0=\sqrt{6 \over 5\pi\alpha_G} M_U$ and
$\alpha_G=g_0^2/4\pi$ is the GUT coupling constant. We now
introduce the parameters,
\be
\epsilon^{(1)}= \biggl\lbrack \biggl\lbrace{1 \over
25\pi\alpha_G} \biggr\rbrace^{1 \over 2} {M_U  \over
M_{Pl}}\biggr\rbrack \eta^{(1)}
\ee
and
\be
\epsilon_i^{(2)}= \biggl\lbrack \biggl\lbrace{1 \over
25\pi\alpha_G} \biggr\rbrace^{1 \over 2} {M_U  \over
M_{Pl}}\biggr\rbrack^2 \eta_i^{(2)}
\ee
Then the $G_{PS}$ invariant effective lagrangian will be
modified by these higher dimensional operators as follows,
\[
-\frac{1}{2}(1+\epsilon_4) \: {\rm Tr} (F_{\mu
\nu}^{(4)}\:F^{(4)\mu \nu}) -\frac{1}{2}(1+\epsilon_{2}) \: {\rm
Tr} (F_{\mu \nu}^{(2L)}\:F^{(2L)\mu \nu})
\]
\be
-\frac{1}{2}(1+\epsilon_{2}) \: {\rm Tr} (F_{\mu
\nu}^{(2R)}\:F^{(2R)\mu \nu}  )
\ee
where, $$\epsilon_4 = \epsilon^{(1)} + \epsilon^{(2)}_a +
\frac{1}{2} \epsilon^{(2)}_b $$ and
$$\epsilon_2 = - \frac{3}{2} \epsilon^{(1)} + \frac{9}{4}
\epsilon^{(2)}_a + \frac{1}{2} \epsilon^{(2)}_b. $$
Then the usual $G_{PS}$ lagrangian can be recovered with the
modified coupling constants,

\begin{eqnarray}
g_4^2(M_U)&=& \bar{g}_4^2(M_U) (1 + \epsilon_4)^{-1}\nonumber\\
g_{2L}^2(M_U)&=& \bar{g}_{2L}^2(M_U) (1 + \epsilon_2)^{-1}\nonumber\\
g_{2R}^2(M_U)&=& \bar{g}_{2R}^2(M_U) (1 + \epsilon_2)^{-1}
\end{eqnarray}
where, $\bar{g}_i$ are the coupling constants in the absence of
the nonrenormalizable terms and $g_i$ are the physical coupling
constants that evolve below $M_U$. Similarly, the physical
gauge fields are defined as, $A_i^{\prime} = A_i \sqrt{ 1 +
\epsilon_i}$.

The $vev$ of $\Sigma$ leaves unbroken a larger symmetry group
than $G_{PS}$, which is $O(6) \otimes O(4)$. The D-parity is
thus unbroken and hence $SU(2)_L$ and $SU(2)_R$ always receive
equal contributions. Furthermore, since overall contributions
to all the gauge groups cannot change the predictions of
$\sin^2 \theta_w$ and $\alpha_s$, the $vev$ of $\Sigma$ can only
contribute to one combination of the couplings, {\it i.e.}
the relative couplings of $SU(4)$ and the $SU(2)$s. For this
reason no matter how many higher dimensional terms we consider,
what contributes to the low energy predictions of $\sin^2
\theta_w$ and $\alpha_s$ is only the combination,
\be
\epsilon =
\epsilon_4 - \epsilon_2.
\ee

If we now assume that the dimension 6 terms $\epsilon_i^{(2)}$
are negligible compared to the dimension 5 terms
$\epsilon^{(1)}$, then we further get, $$\epsilon_4 =
\frac{2}{5} \epsilon \;\;\;\;\; {\rm and} \;\;\;\;\; \epsilon_2
= - \frac{3}{5} \epsilon. $$ As we argued earlier, this does not
reduce the number of parameters in the theory. If we include
the higher dimensional terms, then the allowed region in
$\epsilon$ will be shared by the other $\epsilon^{(n)}$s.

It was pointed out in ref. \cite{alak} that for any choice of the
parameter $\epsilon$ it was not possible to have a consistent
theory with low $M_R$. It was neccessary to make the symmetry
breaking scale $M_I$ very close to $M_U$, so that higher
dimensional operators can introduce another parameter, which
can then allow low $M_R$.

The $vev$ of a 45-plet field $H$ can break the symmetry group
$G_{PS}$ to $G_{LR}$,
\be
\langle H \rangle = \displaystyle \frac{1}{\sqrt{12}\:
i} \:\: H_0 \: \pmatrix{ 0_{33} & 1_{33} & 0_{34} \cr -1_{33} & 0_{33} &
0_{34} \cr 0_{43} & 0_{43} & 0_{44} }
\ee
where, $0_{mn}$ is a $m \times n$ null matrix and $1_{mm}$ is a
$m \times m$ unit matrix. The antisymmetry of the matrix $H$ will
imply that to dimension five operators there is no contribution
from this higgs. The lowest order contribution comes from the
dimension six operators,( In Ref 12.  the dimension
five operator
was taken to give the lowest order contribution.
This is incorrect,since due to the
 antisymmetry
of the $45$ representation the dimension five operator is zero.)
\begin{eqnarray}
{L^\prime}^{(2)}&=&-{1 \over 2} {1 \over M_{Pl}^2} \biggl\lbrack
\eta_a^{\prime(2)} {\rm Tr} (F_{\mu\nu} H^2 F^{\mu\nu})
+ \eta_b^{\prime(2)} {\rm Tr}(H^2){\rm Tr} (F_{\mu\nu} F^{\mu\nu})
\nonumber \\ && +
\eta_c^{\prime(2)} {\rm Tr} (F^{\mu\nu}H) {\rm Tr} (F_{\mu\nu}
H) \biggr\rbrack \label{eqn21}
\end{eqnarray}
The $vev$ of $H$ doesnot modify the $SU(2)$ couplings.
The $SU(4)$ invariant effective lagrangian will only contain a new
contribution,
\begin{eqnarray}
{L^{\prime\prime}}^{(2)}&=&-{1 \over 2} {1 \over M_{Pl}^2} \biggl\lbrack
\eta_a^{\prime(2)} {\rm Tr} (F_{\mu\nu} \phi_{15}^2 F^{\mu\nu})
+ \eta_b^{\prime(2)} {\rm Tr}(\phi_{15}^2){\rm Tr} (F_{\mu\nu} F^{\mu\nu})
\nonumber \\ && +
\eta_c^{\prime(2)} {\rm Tr} (F^{\mu\nu}\phi_{15}) {\rm Tr} (F_{\mu\nu}
\phi_{15}) \biggr\rbrack \label{eqn21}
\end{eqnarray}
where, $\phi_{15}$ transforms as $(15,1,1)$ of $G_{PS}$. At
$M_I$ the symmetry group $SU(4)_c$ breaks down to $SU(3)_c
\otimes U(1)_{B-L}$ when the field $\phi_{15}$ acquires a $vev$,
\be
\phi_{15} = \frac{1}{\sqrt{24}} \phi_0 {\rm diag}
[1, 1, 1, -3].
\ee
with, $\phi_0 = \sqrt{6/5 \pi \alpha_4} M_I$. We now define,
\be
\epsilon_i^{\prime(2)} = \displaystyle \frac{\eta_i^{\prime(2)}
\phi_0^2}{ 24 M_{Pl}^2} = \left[ \frac{1}{20 \pi \alpha_4}
\biggl\lbrack\frac{M_I}{M_{Pl}}\biggr\rbrack^{2} \right] \eta_i^{\prime(2)}
\ee
where, $i =
a, b, c$. The $SU(3)_c
\otimes U(1)_{B-L}$ invariant kinetic energy term for the gauge
bosons will then be given by,
\be
-\frac{1}{2}(1+\epsilon_3^\prime) \: {\rm Tr} (F_{\mu
\nu}^{(3)}\:F^{(3)\mu \nu}) -\frac{1}{2}(1+\epsilon_{1}^\prime)
\: {\rm Tr} (F_{\mu \nu}^{(1)}\:F^{(1)\mu \nu})
\ee
where, $$ \epsilon_3^\prime = \epsilon_a^{\prime(2)} + 12
\epsilon_b^{\prime(2)} $$ and $$ \epsilon_1^\prime =
7 \epsilon_a^{\prime(2)} + 12 \epsilon_b^{\prime(2)} + 12
\epsilon_c^{\prime(2)}. $$
In general, $\epsilon_1^\prime$ and $\epsilon_3^\prime$ may be
treated as two free parameters. But we shall assume $
\epsilon_b^{\prime(2)} = \epsilon_a^{\prime(2)} =
\epsilon_c^{\prime(2)} $ and hence,
\be
\epsilon_3^\prime =
0.42 \;\;\; \epsilon_1^\prime = \epsilon^\prime ({\rm say}).
\ee
Thus the parameter space in $\epsilon^\prime$ and $\epsilon$ we
present here may be further relaxed to some extent. However,
the number of parameters in the $\sin^2 \theta_w$ and
$\alpha_s$ is not changed and we cannot expect any change in
low energy predictions. In our analysis we shall present the
parameter space of $\epsilon^\prime$ and $\epsilon$, which
allows low $M_R$.For the D-nonconserved case a $210$ plet of higgs
is used to break $SO(10)$ to the group $G_{PS}$ without D-parity
conservation.The $vev$ of the Higgs is given by,
\begin{eqnarray}
\langle H_{210}\rangle = \displaystyle
\frac{1}{\sqrt{32}}\:\: H_{0} \:\: {\rm diag}(1_{44},1_{44},-1_{44},-1_{44}).
\end{eqnarray}
where $H_0$ is related to the vector boson mass $M_X$ by $\sqrt{\frac{2}{\pi
\alpha_{G}}}M_{X}=H_{0}$.
Keeping only the dim-5 operator we get,
$$\epsilon_{4}=0 $$
 $$\epsilon_{2L}=-\epsilon_{2R}= 8 \epsilon^{(1)}=\epsilon $$
where,
$$\epsilon^{(1)} =\sqrt{\frac{2}{32\pi\alpha_{G}}}
\biggl\lbrack\frac{M_{X}}{M_{Pl}}\biggr\rbrack$$

For the evolution of the coupling constants we use the two loop
renormalization group equations \cite{amaldi,palas,mohpar},
\be
\mu \frac{ \partial}{\partial \mu} \alpha_i(\mu) = \frac{2}{4
\pi} \left[ b_i + \sum_j \frac{b_{ij}}{4 \pi} \alpha_j(\mu)
\right] \alpha_i^2 (\mu) + \frac{2 b_{ij}}{(4 \pi)^2}
\alpha_i^3
\ee
where $i,j$ index represents the different subgroups at the
energy scale $\mu$ and $\alpha_i = \frac{1}{4 \pi} g_i^2.$
The various $\beta$-functions with SUSY and without SUSY are given in Ref 19
 \cite{Jones}.
We use the suvival hypothesis\cite{SH} to find the Higgs content at
the various mass scales for any given chain.In  table 1. we
show the higgs bosons that live at different mass scales.
\begin{table}
\caption{Higgs spectrum at various mass scales for the D-conserved and
the D-nonconseved chain}
\begin{center}
\begin{tabular}{|c|c|}
\hline
$ Group$ $ G_i $ & $ Higgs$ $ content$
 \\
\hline
                    & $ (2,2,1)_{10} $ \\
$ (2_{L}2_{R}4_{C} P) $ & $ (1,3,{\bar{10}})_{126} $ \\
                      & $ (3,1,10)_{126} $ \\
                      & $ (1,1,15)_{45} $ \\
\hline
                     & $ (2,2,1)_{10} $ \\
$ (2_{L}2_{R}4_{C}) $ & $ (1,3,{\bar{10}})_{126} $ \\
                      & $ (1,1,15)_{45} $ \\
\hline
                      & $ (2,2,0,1)_{10} $ \\
$ (2_{L}2_{R}1_{B-L}3_{C}P) $ & $ (1,3,2,1)_{126} $ \\
                             & $ ((3,1,2,1)_{126} $ \\
\hline
                         & $ (2,2,0,1)_{10} $ \\
$ (2_{L}2_{R}1_{B-L}3_{C}) $ & $ (1,3,2,1)_{126} $ \\
\hline
\end{tabular}
\end{center}
\end{table}

An approximate solution of the evolution  equation can be written as,
\begin{eqnarray}
\alpha_i^{-1}(\mu^\prime) & = & \beta_0 \ln \frac{\mu^\prime}{\mu} +
\frac{1}{\alpha_i (\mu)} + \frac{\beta_1}{\beta_0} \ln \left[
\displaystyle \frac{\alpha^{-1}(\mu^\prime) +
\frac{\beta_1}{\beta_0} }{\alpha^{-1}(\mu) +
\frac{\beta_1}{\beta_0} } \right] \nonumber \\
\beta_0 & = & \frac{1}{2 \pi} \left[ b_i + \sum b_{ij} \alpha_j
(\mu) \right] \nonumber \\
\beta_1 & = & - \frac{2 b_{ii}}{(4 \pi)^2}
\end{eqnarray}
At each symmetry breaking threshold we use the following matching conditions
for the couplings when the group $G$ breaks to the group $G_i$ \cite{wl}
\begin{eqnarray}
\alpha_i^{-1}(\mu) & = & \alpha_G^{-1} - \frac{\lambda_i}{12\pi}
\end{eqnarray}
where,
$$\lambda_i  =  C_G - C_G{}_i +{\rm Tr}(\theta_i^{H})^{2}\ln\frac{M_H}{\mu}$$

 $\theta_i^{H}$ are the generators of $G_i$ for the representation
in which the higgs,$M_H$, appear.$C_G$ and $C_G{}_i$ are the quadratic Casimir
invariants for the group $G$ and the group $G_i$ while $\mu$ is the symmetry
breaking scale.Gravity induced corrections change $\alpha_i^{-1}(\mu)$ to
$\alpha_i^{-1}(\mu) (1+\epsilon_i)^{-1}$ as in eqn.(11).
In our analysis
we identify $\mu$,the unification scale with the vector boson mass.Threshold
corrections
will occur due to the non-degeneracy of the higgs masses with the vector
boson mass.Using the D-conserved non-susy case as an example we have calculated
the effect of threshold corrections at the heavy scales $M_I$ and $M_U$.The
higgs
 masses  are assumed to vary between $\frac{1}{5}$ and $5$ times the vector
boson mass.The threshold corrections enter through the factors $\lambda_{i}$
appearing in eqn.24.The higgs that live around the mass scale $M_U$ and
$M_{I}$ are given in table 3. and table 4.
\begin{table}
\caption{The Higgs bosons at $M_U$ }
\begin{center}
\begin{tabular}{|c|c|}
\hline
$SO(10)$ $multiplet$ & $G_{2,2,4}$ $multiplet$\\
\hline
54 & $S_1(1,1,1)$,$S_2(1,1,20^{\prime})$,$S_{3}(3,3,1)$\\
\hline
45 & $\phi_{1}(3,11)$,$\phi_{2}(2,2,6)$,$\phi_{3}(1,3,1)$\\
\hline
126 & $\Sigma_{1}(2,2,15)$,$\Sigma_{2}(1,1,6)$\\
\hline
\end{tabular}
\end{center}
\end{table}
\begin{table}
\caption{The Higgs bosons at $M_I$ }
\begin{center}
\begin{tabular}{|c|c|}
\hline
$SO(10)$ $multiplet$ & $G_{2_{L},2_{R},1_{B-L},3_{C}}$ $multiplet$\\
\hline
45 & $Y_{1}(1,1,1,0)$,$Y_{2}(1,1,0,8)$,$\bar{X}(1,1,\frac{4}{3},\bar{3})
$,$X(1,1,-\frac{4}{3},3)$\\
\hline
126 & $\xi_{1}(3,1,-\frac{2}{3},6)$,$\xi_{2}(1,3,\frac{2}{3},\bar{6})$,
$\xi_{3}(3,1,\frac{2}{3},3)$,$\xi_{4}(1,3,-\frac{2}{3},\bar{3})$\\
\hline
\end{tabular}
\end{center}
\end{table}

Defining $\eta_{H}=\ln\frac{M_{H}}{M_{X}}$ one can write at $M_{U}$,
\begin{eqnarray}
\lambda_{4} &= &(4+16\eta_{S_{2}} +8\eta_{\phi_{2}} +32\eta_{\Sigma_{1}}
+2\eta_{\Sigma_{2}} +2\eta_{H^{\prime}})\nonumber\\
\lambda_{2}=&= &(6+12\eta_{S_{3}} +12\eta_{\phi_{2}} +30\eta_{\Sigma_{1}}
 +4\eta_{\phi_{1}})
\end{eqnarray}
At the scale $M_{I}$ one has,
\begin{eqnarray}
\lambda_{3} & = & (1+6\eta_{Y_{2}} +15\eta_{\xi_{1}} +15\eta_{\xi_{2}}
+3\eta_{\xi_{3}} +3\eta_{\xi_{4}})\nonumber\\
\lambda_{B-L} & = & (4+ 6\eta_{\xi_{1}} +6\eta_{\xi_{2}}
+3\eta_{\xi_{3}} +3\eta_{\xi_{4}})\nonumber\\
\lambda_{2L} & = & (24\eta_{\xi_{1}} + 12\eta_{\xi_{3}})
\nonumber\\
\lambda_{2R} & = & (24\eta_{\xi_{2}} + 12\eta_{\xi_{4}})
\end{eqnarray}
The quantities $\lambda_4 - \lambda_2 $ and $\lambda_{3} -\lambda_{B-L}$
appear in the solution for $M_{U}$ and $M_{I}$ respectively.We consider two
cases
where the higgs masses are choosen such that the above quantities are at their
extreme values.We further make the assumption that the higgs at a given scale
coming from the same SO(10) multiplet have the same masses.
 In the first case we choose $M_{\Sigma}$,$M_{S
}$,$M_{H^{\prime}}$ to be $(\frac{1}{5})M_{U}$ while $M_{\phi}$ to be $5M_{U}$.
At $M_{I}$ we choose $M_{Y}$ and $M_{\xi}$ to be $\frac{1}{5}M_{I}$.For the
second case we just flip the higgs bosons around at the two scales.
We refer to these two cases as case(a) and case(b).We have
only considered the cases when $M_I$ is not equal to $M_U$.
\subsection{Results}
Using the values of the standard model couplings at $M_Z$ (eqn.1),the evolution
equations and the matching conditions (eqn.24 and eqn.25) we find  regions in
the $ \epsilon$ , $\epsilon^{\prime}$ space which allow a low $M_R$ for various
values of the intermediate scale $M_I$ and the unification scale $M_U$.
In fig 1. the allowed regions for D-conserved non-susy case are shown.
For $M_U$ not equal to $M_I$ the effects of threshold corrections have
been included. In fig 2. and fig 3. the allowed regions for D-conserved
 and D-non-conserved susy case are shown.
For the D-broken non-susy case the width of the allowed regions
are
too small to be shown graphically and therefore we present the results for this
case in table 4.For the supersymmetric version the allowed regions are larger
but
no solution was found for the case $M_I=10^{16}$,$M_U=10^{18}$.
 Even though we
have not carried out a full analysis of the threshold effects, from the
examples
considered, we do not expect moderate threshold effects to alter the regions in
the parameter space drastically.In conclusion we have shown that both for
the D-conserved and D-nonconserved case we can find regions in the parameter
space of gravity induced couplings that allow  $M_{R}$ in the {\rm TeV}
range.

\begin{table}
\caption{Allowed ranges for $\epsilon $ and $\epsilon^{\prime}$ for D-broken
non-susy case}
\begin{center}
\begin{tabular}{|c|c|c|c|c|}
\hline
$\epsilon^{\prime}$ $(10^{-3})$ & $ \epsilon $ $(10^{-3})$
 &$ M_I$ & $M_U$ & $\frac{\alpha_{2L}(M_R)}
{\alpha_{2R}(M_R)} $ \\
\hline
4.92 -- 10.38 & -10 -- -5.43 & $ 10^{16} $ & $ 10^{17} $ & 1.4 \\
\hline
11.69 -- 12.46 & -9.72 -- -1.032 & $ 10^{16}$ & $10^{18}$ & 1.3 \\
\hline
6.46 -- 8.46 & -7.09 --  -5.32 & $10^{17}$ & $10^{17}$ & 1.3 \\
\hline
11.08 -- 13.23 & -9.59 -- -7.91 & $10^{17}$ & $10^{18}$ & 1.2 \\
\hline
9.23 -- 10.92 & -9.18 -- -7.75 &  $10^{17}$ & $10^{18}$ & 1.3 \\
\hline
8.23 -- 10.4 & -7.33 -- -5.50 & $10^{18}$ & $10^{18}$ & 1.2 \\
\hline
\end{tabular}
\end{center}
\end{table}
\subsection{Acknowledgement}
One of us (US) would like to acknowedge the hospitality of the
High Energy Physics Group of the University of Hawaii where
this work was initiated and one (S.P) would like to thank the Physical
Research Laboratory for their hospitality.
 This work was supported in part  by US
D.O.E under contract DE-AMO3-76SF-00325.

\newpage

\subsection{Figure Captions}
\begin{itemize}
\item[{\bf Fig. 1:}] The allowed regions in $\epsilon$ and $\epsilon^{\prime}$
space
for D-conserved non-susy SO(10)
for pairs of $M_I$ and $M_U$.For $M_I$ not equal to $M_U$ the upper and
the lower regions correspond to case (a) and case (b),the two cases considered
for threshold corrections.

\item[{\bf Fig. 2:}] The allowed regions in $\epsilon$ and $\epsilon^{\prime}$
space for D-conserved SO(10).

\item[{\bf Fig. 3:}] The allowed regions in $\epsilon$ and $\epsilon^{\prime}$
space
for D-nonconserved SO(10).
\end{itemize}

\end{document}